\documentclass[preprint]{revtex4}

\usepackage{amsmath,amssymb}
\usepackage{graphicx}
\usepackage{mhchem}

\begin{document}
\title{From the ``Brazuca'' ball  to  Octahedral Fullerenes: 
Their Construction and Classification
}
\author{Yuan-Jia Fan}
\author{Bih-Yaw Jin} 
\email{byjin@ntu.edu.tw}
\affiliation{Department of Chemistry and Center for Emerging Material
  and Advanced Devices, National Taiwan University, Taipei 10617,
  Taiwan}

 \date{\today}
\begin{abstract}
  A simple cut-and-patch method is presented for the construction and
  classification for fullerenes belonging to the octahedral point
  groups, $O$ or $O_h$.  In order to satisfy the symmetry requirement
  of the octahedral group, suitable numbers of four- and eight-member
  rings, in addition to the hexagons and pentagons, have to be
  introduced.  An index consisting of four integers is introduced to
  specify an octahedral fullerenes. However, to specify an octahedral
  fullerene uniquely, we also found certain symmetry rules for these
  indices. Based on the transformation properties under the symmetry
  operations that an octahedral fullerene belongs to, we can identify
  four  structural types of octahedral fullerenes.
\end{abstract}
\maketitle

\section{Introduction}
\label{sec:introduction}
The 2014 FIFA World Cup championship in Brazil expected to
attract the attention of more than 3 billion people worldwide is the
quadrennial soccer tournament since 1930.  Whereas, the most popular
design for the FIFA soccer ball, the Telstar, which was used in the
official logo for the 1970 World Cup, consists of twelve black pentagonal
and 20 white hexagonal panels, a truncated icosahedron belonging
icosahedral point group\cite{soccer_ball,kotshick:05}. Since then, a
number of different designs have appeared, some with small variations
such as the Fevernova (2002) and the Teamgeist (2006) which still have
icosahedral symmetry but with low-symmetry tetrahedral patterns
painted on the ball; some balls only show lower polyhedral patterns
such as Jabulani (2012) with tetrahedral symmetry without icosahedral
symmetry superimposed.  However, the soccer ball, ``Brazuca'' ball,
used for the World Cup this summer in Brazil has a new design based on
octahedral symmetry.  Basically, the ``Brazuca'' ball is composed of
six bonded polyerethane panels with four-arm clover-shaped panels that
interlock like a jigsaw puzzle smoothly on a
sphere\cite{brazuca,asai:14}.

%This structure of carbon allotropes form a completely spherical shape
%which is called buckyball which belong to a more widely class of pure
%sp$^2$ carbon allotropes!

In 1985, it was discovered that, in addition to diamond and
graphite, carbon atoms can have a third new allotrope consisting of
60-atom spherical molecules, $\ce{C60}$, sometimes nicknamed molecular
soccer ball because the shape of this molecule is identical to the
standard soccer ball, with $60$ atoms located at $60$ identical
vertices\cite{Smalley85}. More generally, this molecule belongs to a
family of sp$^2$-hybridized pure carbon systems now called fullerenes
that contain only five- and six-membered rings.  Since then,
structures of fullerenes have been extensively studied experimentally
and theoretically.  Under this constraint, considerable effort has
been devoted to detailed enumerations of possible structures. For
instance, a complete list of fullerenes with less than or equal to 60
carbon atoms and all fullerenes less than and equal to $100$ carbon
atoms that satisfy the isolated pentagon rule (IPR) is tabulated in
the monograph by Fowler and Manolopolous\cite{Fowler07}. Among all
these fullerenes $C_{N}$, $N\le 100$, the possible symmetry point
groups for fullerenes are $C_1$, $C_s$, $C_i$, $C_m$, $C_{mv}$,
$C_{mh}$, $S_{2m}$, $D_n$, $D_{nd}$, $D_{nh}$, $T$, $T_d$, $T_h$, $I$
and $I_h$, where $m$ can be $2$ or $3$ and $n$ can be $2$, $3$, $5$ or
$6$. However, only two out of three Platonic polyhedral groups, namely
tetrahedral and icosahedral groups, seems to be possible for
fullerenes. So the question is, can we have fullerenes with octahedral
symmetry just like the ``Brazuca'' ball? If possible, what are the
general construction and classification rules for this family of
octahedral fullerenes?
 
\begin{figure}[h]
  \centering
  \includegraphics[width=12cm]{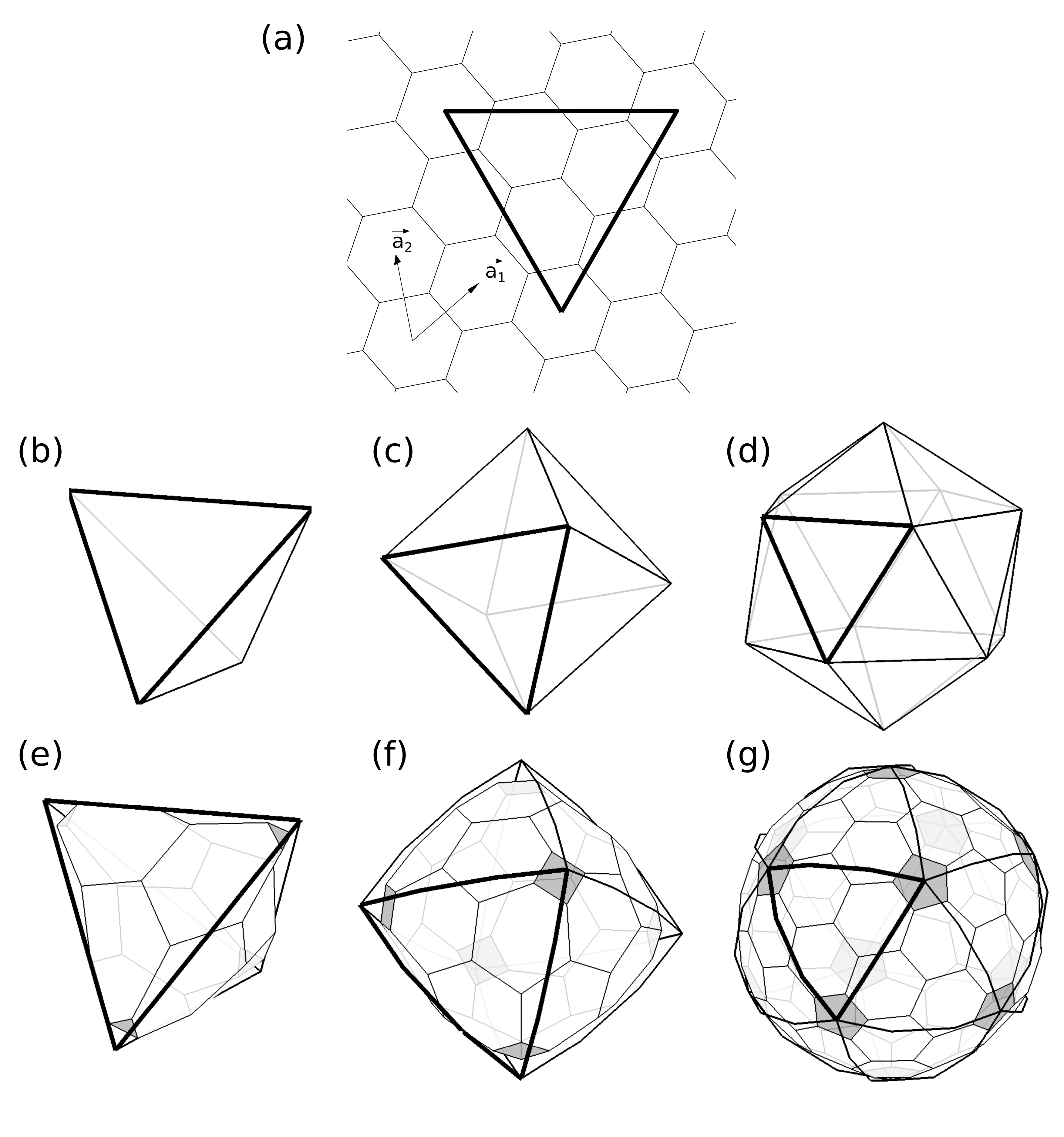}
  \caption{Goldberg polyhedra by the cut-and-patch construction. Here we
    cut a equilateral triangle which can be specified by an vector
    $(2,1)$ (also known as Goldberg vector) from graphene
    and then patch the triangle onto different platonic solids to
    construct fullerenes with different polyhedral symmetries. The
    famous $\ce{C60}$, can be also constructed in this way using
    icosahedron as the template with Goldberg vector (1,1). 
    \label{Fig:Goldberg}}
\end{figure}

To answer this question, we start with the construction process of
fullerenes with polyhedral symmetries through a simple cut-and-patch
procedure as shown in Figure~\ref{Fig:Goldberg}. For instance,
constructing a fullerene with icosahedral symmetry can be done by
cutting 20 equivalent equilateral triangles from graphene and pasting
them onto the triangular faces of an icosahedron. This will create
twelve pentagons sitting at twelve vertices of the
icosahedron\cite{Goldberg37,Caspar62,Fowler92}. Similar cut-and-patch
procedure can be used to construct fullerenes with tetrahedral and
octahedral symmetries, too (Figure~\ref{Fig:Goldberg}). However, the
non-hexagons such as triangles and squares will appear at the vertices
of the template tetrahedron and octahedron, which are in contradiction
to the definition of fullerenes. In the case of tetrahedral
fullerenes, we can replace the template tetrahedron with a truncated
tetrahedron.  This makes it possible to the construction of
tetrahedral fullerenes without triangles by a suitable cut-and-patch
construction scheme\cite{Fowler88}.  But this technique is not
applicable to octahedral fullerenes\cite{Fowler93,Kardos07}.

Albeit the appearance of squares in these caged octahedral fullerenes
leads to energetically unstable molecules, one can still find in
literatures that some studies have been carried out on the geometric,
topological\cite{Fowler01}, and electronic
structures\cite{Huang96,Ceulemans05,Ceulemans05_2,Dunlap94} of
fullerenes with octahedral symmetry by introducing squares on a
template octahedron (Figure~\ref{Fig:Goldberg}).  In addition to the
pure carbon allotropes, the octahedral boron-nitride systems have also
been vigorously
investigated\cite{Jiao04,Scuseria06,Rogers00,Benson00,Dunlap04}.

In this paper, we present a general cut-and-patch construction and
classification scheme for fullerenes with octahedral symmetry by
systematically introducing some other non-hexagons such as octagons with a
cantellated cube as the template. The octahedral fullerenes previously considered in
literatures are included as limiting cases in our general construction
scheme\cite{Fowler01,Huang96,Ceulemans05,Ceulemans05_2,Dunlap94}. We
also like to point out that the cut-and-patch method is a simple and
powerful method for building various kinds of fullerenes and graphitic
structures.  For instance, we have applied this method successfully to
many other template polyhedral tori and concluded general structural
rules of carbon nanotori\cite{Chuang09_1,Chuang09_3}. From there,
structural relations for a whole family of topologically nontrivial
fullerenes and graphitic structures such as carbon nanohelices,
high-genus fullerenes, carbon Schwarzites and so on can be
derived\cite{Chuang09_2, jin:2013,jin:2010x,jin:2011a}.

\section{Requirement of Octahedral Fullerenes}

We start by briefly describing the icosahedral fullerenes that consist
only of hexagons and pentagons.  The simplest icosahedral fullerene
that satisfies IPR is $\ce{C60}$, which can also be viewed as a
truncated icosahedron, one of the thirteen Archimedean solids if we
ignore the slight variation in bond lengths. In a truncated
icosahedron, there are exactly twelve pentagons and twenty
hexagons. This structure can be derived from a regular icosahedron by
truncating the twelve vertices away appropriately. We will show that this is the
only possibility if we want to construct an icosahedral fullerene with
pentagons and hexagons only.

Using the Euler's polyhedron formula, $V-E+F=2$ for a polyhedron with
$V$ vertices, $E$ edges and $F$ faces, and the condition $3V=2E$ for
trivalent carbon atoms in a fullerene, we can find easily the
condition $\sum_{n}(6-n)F_{n}=12$, where $F_{n}$ is the number of
$n$-gons. If we assign each face a topological charge $6-n$, the
Euler's polyhedron formula states that the sum of topological charges
of a trivalent polyhedron must be twelve.  Therefore, fullerenes that
contain only pentagons and hexagons must have twelve pentagons,
i.e. $F_{5}=12$, while there is no constraint on the number of
hexagons, $F_{6}$, except the case with only one hexagon,
$F_6=1$, is forbidden. This conclusion is general and can be applied to any
fullerene regardless of its symmetry.

An arbitrary icosahedral fullerene can be classified by its chiral
vector $(h, k)$, where $h$ and $k$ satisfy the inequality $h\ge k\ge
0\land h>0$, according to the Goldberg construction
\cite{Goldberg37}. For instance, $\ce{C60}$ corresponds to the
fullerene with chiral vector $(1,1)$. Interestingly, these twelve
pentagons are located at the high-symmetry points along the six fivefold
rotational axes of the icosahedral symmetry group.  Suppose that these
pentagons are not located at the high symmetry points, there should be
five pentagons around each of these points in order to satisfy the
symmetry requirement. Therefore, there must be $12\times 5 =
60$ pentagons in total. However, the condition, $\sum_n (6-n)F_n=12$,
will require some $n$-gons where $n>6$ to compensate the extra
topological charges introduced by these pentagons.
So, we conclude that exactly twelve pentagons must be located at the high symmetry
points along the six fivefold rotation axes.

We apply the above analysis to the requirement for
octahedral fullerenes. First, there are three fourfold axes, four
threefold axes, and six twofold axes in the octahedral group.
Pentagons are not compatible with any of the high symmetry points of
octahedral groups. Therefore, there is no high symmetry point where
pentagons can be located. The best we can do is to put clusters of
pentagons around, for example, the three fourfold axes. Then we need to
put twenty-four pentagons together with $n$-gons where $n>6$ to balance the
topological charges. A simple way is to put six octagons at the six
high symmetry points along three fourfold axes, so that the condition,
$\sum_{n}(6-n)F_{n}=12$, is satisfied.  The twofold or the fourfold axes can
also be chosen\cite{Kardos07}, but the resulting fullerenes are
considerably more energetically unfavored because additional non-hexagons
need to be introduced.

\begin{figure}
\includegraphics[width=15cm]{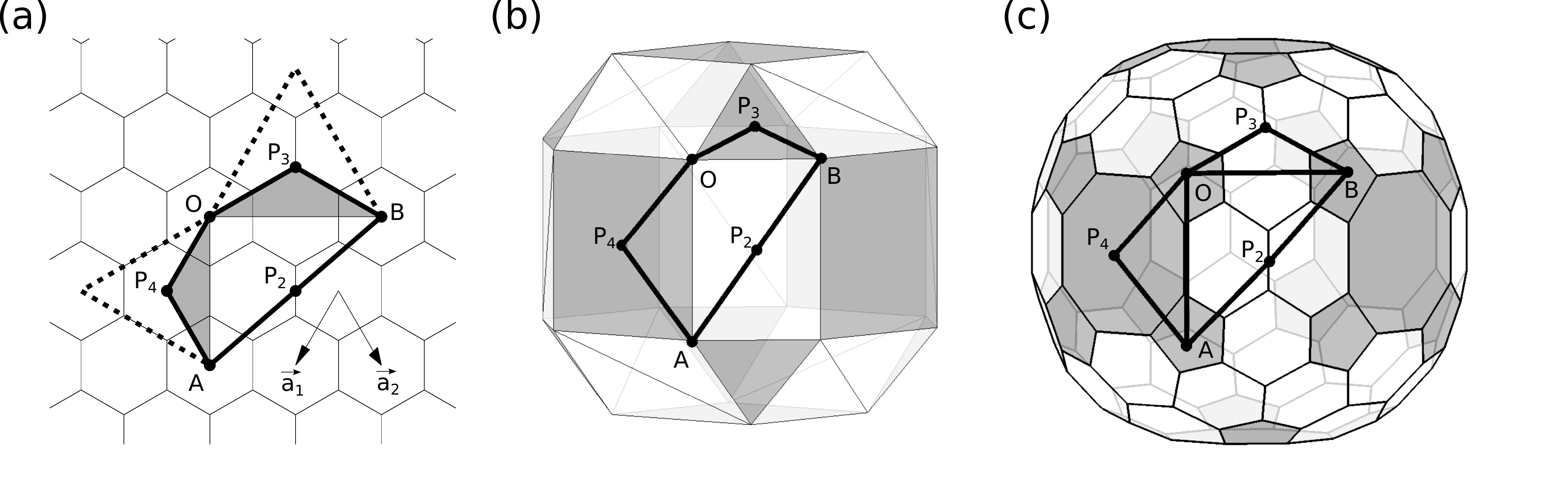}
\caption{Cut-and-patch procedure for constructing an octahedral
  fullerene.  
%These three figures show how a cut-and-patch procedure is done. 
  Points, $P_{2}$, $P_{3}$ and $P_{4}$, represent the high symmetry
  points of the octahedral symmetry respectively. $\bigtriangleup
  OP_{4}A$ ($\bigtriangleup OP_{3}B$) is one-third of a regular
  triangle (the dotted triangle in (a)) and $P_{4}$ ($P_{3}$) is the
  corresponding triangle center on graphene. Points $O$,
  $A$, and $B$ become positions where twenty-four equivalent pentagons
  are located at, while point $P_{4}$ becomes the position for one of six
  equivalent octagons after they are patched onto the cantellated
  cube.  The two base vectors, $\protect\overrightarrow{OA}=(i,j)$ and
  $\protect\overrightarrow{OB}=(k,l)$, can in general be any two
vectors such that $P_{4}$ does not coincide with an atom
  ({\it i.e.}, $i-j=3n$). $\{i,j,k,l\}=\{1,1,-2,2\}$ in this example. (c)
  The 3D geometry of an octahedral fullerene specified according to
  its topological coordinates \cite{Fowler92}.
\label{fig:OF_Cut_Patch_Scheme}}
\end{figure}

To illustrate this idea, we present a simple construction procedure
using the cut-and-patch scheme as shown in
Figure~\ref{fig:OF_Cut_Patch_Scheme}.  We first cut the polygonal
region as defined by the solid thick line from graphene
(Figure~\ref{fig:OF_Cut_Patch_Scheme}(a)) and then patch twenty-four replica
of it on a cantellated cube as shown in
Figure~\ref{fig:OF_Cut_Patch_Scheme}(b). The points, $O$, $A$, and $B$
in Figure~\ref{fig:OF_Cut_Patch_Scheme}(a) overlap with vertices of
the cantellated cube while $P_3$ and $P_4$ the centers of the triangle
and the square faces, respectively.  In this process, every four
identical isosceles triangles, $\bigtriangleup OP_{4}A$, cover one
square face (Figure~\ref{fig:OF_Cut_Patch_Scheme}(b)).  Since the
angular deficit at $P_{4}$ is $2\pi-4\times 2\pi/3=-2\pi/3$, it must
correspond to the location of an octagon.  On the other hand, the
angular deficit at $O$ is $2\pi-(\pi+2\pi/3)=\pi/3$.  Therefore a
pentagon will be generated at $O$ by this cut-and-patch process.
 
We will define the area inside the solid thick line as shown in
Figure~\ref{fig:OF_Cut_Patch_Scheme}(a) as the fundamental polygon. Note that the two base vectors, $\overrightarrow{OA}=(i,j)$
and $\overrightarrow{OB}=(k,l)$, in the fundamental polygon become the
edges of the square and the regular triangle on the cantellated cube,
respectively, as shown in Figure~\ref{fig:OF_Cut_Patch_Scheme}(b). For
convenience, we refer to $(i,j)$ as the square base vector and $(k,l)$
the triangular base vector from now on.  Using these two vectors, we
can uniquely specify a scalene triangle with four integers
$\{i,j,k,l\}$, which we will simply call the indices of octahedral
fullerenes later. In additional to this scalene triangle, we also need
to incorporate two extra triangles, $\bigtriangleup OP_{4}A$ and
$\bigtriangleup OP_{3}B$, corresponding to one-third of the regular
triangles which share the same edges with the scalene triangle.  The
numbers of carbon atoms inside $\bigtriangleup OP_{4}A$,
$\bigtriangleup OP_{4}A$, and $\bigtriangleup OAB$ are
$(i^2+ij+j^2)/3$, $(k^2+kl+l^2)/3$, and $|il-jk|$, respectively. After
patching twenty-four fundamental polygons onto a cantellated cube, we
get an octahedral fullerene with $8(i^2+ij+j^2+k^2+kl+l^2)+24|il-jk|$
carbon atoms.

\begin{figure}
\includegraphics[width=15cm]{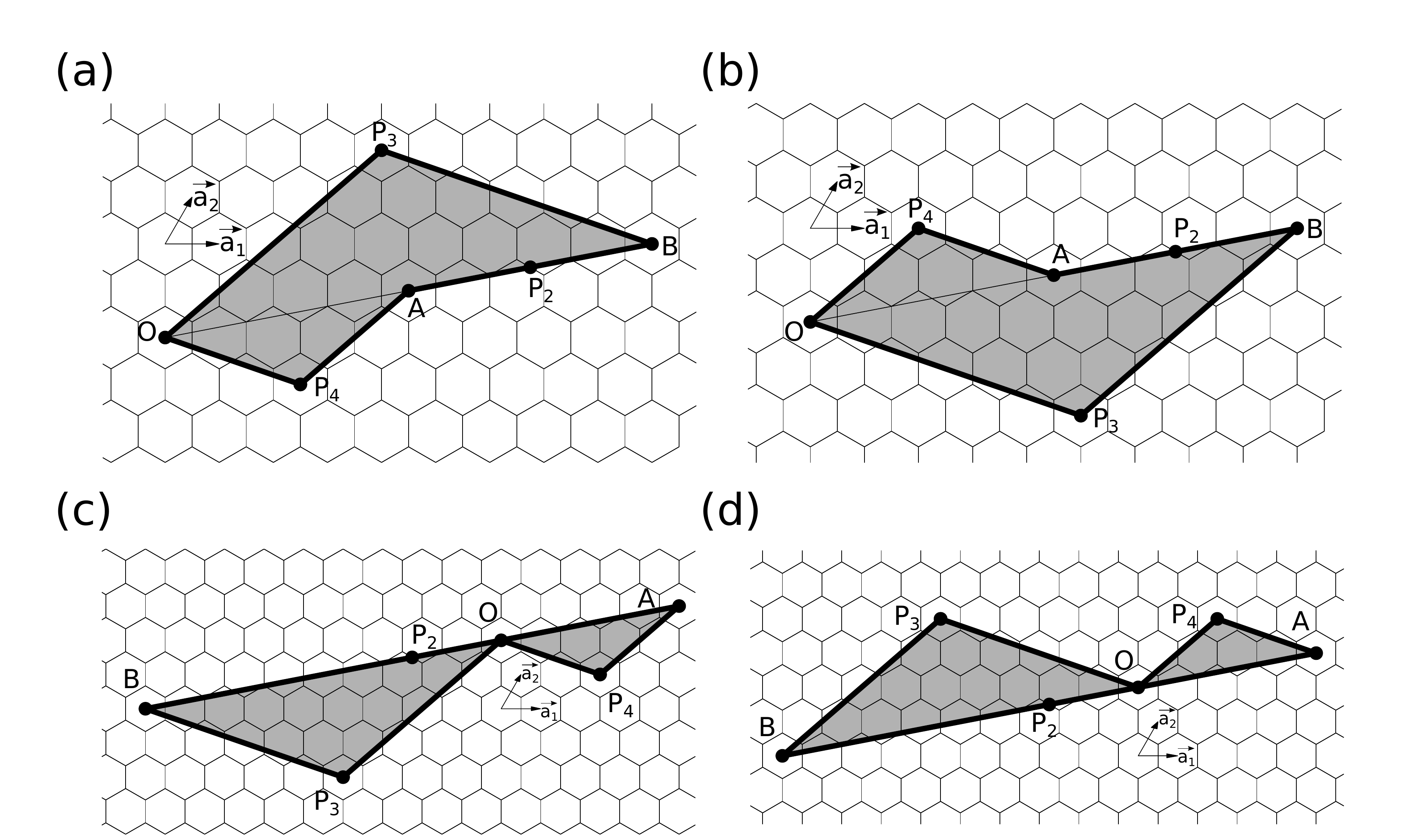}
\caption{Four degenerate cases of octahedral fullerenes. (a)
  $\{4,1,8,2\}_{\alpha}$, (b) $\{4,1,8,2\}_{\beta}$, (c)
  $\{4,1,-8,-2\}_{\alpha}$ and (d) $\{4,1,-8,-2\}_{\beta}$. We have
  $\{4,1,8,2\}_{\alpha}\neq\{4,1,8,2\}_{\beta}$ and
  $\{4,1,-8,-2\}_{\alpha}\neq\{4,1,-8,-2\}_{\beta}$. However,  
  $T_2\{4,1,8,2\}_{\alpha}=\{4,1,-8,-2\}_{\beta}$ and
  $T_2\{4,1,8,2\}_{\beta}=\{4,1,-8,-2\}_{\alpha}$. The $T_2$
  transformation will be discussed in later
  sections. 
  \label{fig:Degenerate}}
\end{figure}

%   === now here
The octahedral fullerenes can be catagorized into two groups according
to the sign of the angle $\theta$ formed by $\overrightarrow{OA}$ and
$\overrightarrow{OB}$.  Octahedral fullerenes with $\pi>\theta>0$ are
in category $\alpha$, $\{i,j,k,l\}_{\alpha}$, and octahedral
fullerenes $-\pi<\theta<0$ are in category $\beta$,
$\{i,j,k,l\}_{\beta}$.  This criterion is equivalent to determining
the sign of $il-jk$, which stands for the signed area enclosed by the
parallelogram spanned by the two base vectors up to a positive
factor. Here we can take one step further to include the degenerate
cases, {\it i.e.} when $\bigtriangleup OAB$ degenerates into a line, which
can be considered as limiting cases when $\theta$ approaches to the
boundaries of its range in each category. It is worthwhile to note
that in general $\lim_{\theta\to0^+}\{i,j,k,l\}_{\alpha}$ is
inequivalent to $\lim_{\theta\to0^-}\{i,j,k,l\}_{\beta}$ and
$\lim_{\theta\to\pi^-}\{i,j,k,l\}_{\alpha}$ is inequivalent to
$\lim_{\theta\to\pi^+}\{i,j,k,l\}_{\beta}$, as shown in
Figure~\ref{fig:Degenerate}. On the other hand, the category letter in
the subscript can be omitted when there is no ambiguity. We will
elaborate in later sections.

Following the above cut-and-patch scheme, we can define a scalene
triangle and thus the fundamental polygon, given the two base vectors
$(i,j)$ and $(k,l)$ that satisfy the condition, $i-j=3n$.  Each of
these fundamental polygons uniquely defines an octahedral fullerene in
non-degenerate case. When the two base vectors are parallel to each
other, it is necessary to further specify the category explicitly.  It
is worthwhile to note that if the condition $i-j=3n$ is not satisfied,
$P_4$ will coincide with a carbon atom, which is not allowed because this
implies that the carbon atom is tetravalent.  At first sight, one
might think that there exists a one-to-one correspondence between an
index, $\{i,j,k,l\}_{X}$, and an octahedral fullerene. But this
is not true since it is possible that the octahedral fullerenes built
from two different scalene triangles are in fact identical.  We will
study this issue in details in the next section.

\begin{figure}[h]
\includegraphics[width=15cm]{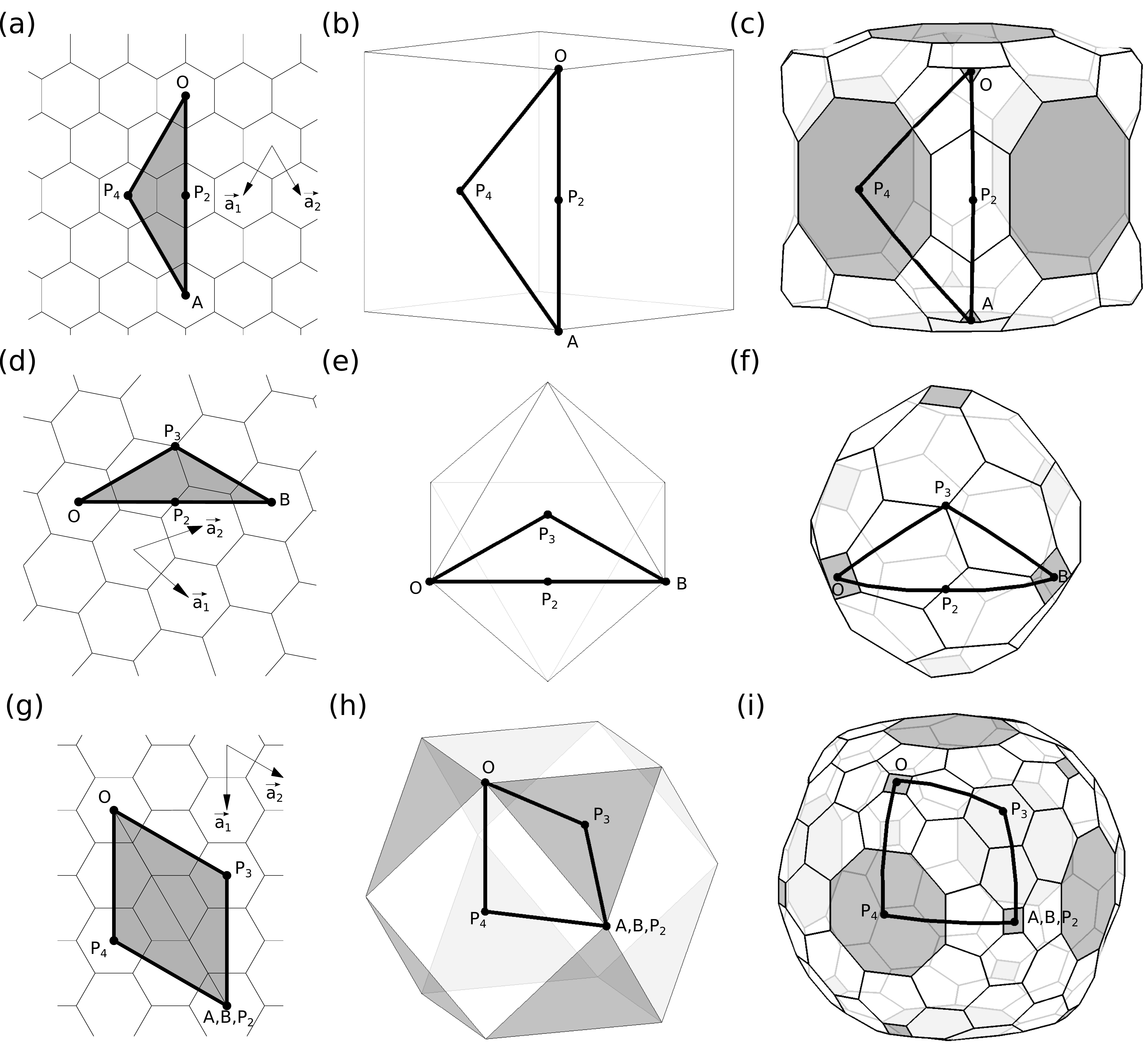}
  \centering
  
  \caption{Three limiting cases of octahedral fullerenes. (a)-(c) Type I
    octahedral fullerene with with $\{2, 2, 0, 0\}$; (d)-(f) Type II
    octahedral fullerene with with $\{0, 0, 1, 2\}$; (g)-(i) Type III
    octahedral fullerene with $\{2, 2, 2, 2\}$. In this case points $A$, $B$, and $P_2$ are
    coincident.} 
  \label{fig:limiting_cases} 
\end{figure}

Finally we can identify three limiting situations if one of the
three sides of the scalene triangle vanishes (see Fig.~\ref{fig:limiting_cases}).
\begin{enumerate}
\item The first limiting situation corresponds to a vanishing
  triangular base vector, $(k,l)=(0,0)$, which is referred to as type
  I octahedral fullerenes later on. The indices for this case have the
  form $\{i,j,0,0\}$.  Thus, the length of the triangular base vector
  $\overrightarrow{OB}$ vanishes and all triangles in the cantellated
  cube shrink to single points. And the template polyhedron reaches
  the corresponding limit of the cantellation, namely the cube. Note
  also that three pentagons fuse to form a triangle at each corner of
  the cube, while the octagons remain at the centers of the faces of
  the cube. Thus, there are eight triangles and six octagons in the
  resulting octahedral fullerene.
\item The second limiting situation corresponds to a vanishing square
  base vector, $(i,j)=(0,0)$, which we denote as type II.  The indices
  for type II fullerenes are given by $\{0,0,k,l\}$. In this limit,
  the length of the square base vector $\overrightarrow{OB}$ vanishes
  and each square shrinks to a point. Thus, the template polyhedron
  reaches another limit of the cantellation, namely the
  octahedron. This case is identical to the Goldberg polyhedron
  illustrated in Figure~\ref{Fig:Goldberg}(c) and
  Figure~\ref{Fig:Goldberg}(f) .  Four pentagons and one octagon fuse
  to form a square at each corner of the octahedron. Therefore, we
  have six squares in a type II octahedral fullerene.
\item The last limiting situation, denoted as type III, is when the
  length of the third side of $\bigtriangleup OAB$,
  $\overrightarrow{AB}$, vanishes. In other words,
  $\overrightarrow{OA}$ is equal to $\overrightarrow{OB}$,
  i.e. $(i,j)=(k,l)$.  One can show that $(i,j)=-(k,l)$ also
  corresponds to the same limiting case. $\{i,j,i,j\}$ and
  $\{i,j,-i,-j\}$ can be transformed to each other via additional
  symmetry transformations, $T_3$ or $T_4$, which will be introduced
  in the next section.  
 The indices for this type are $\{i,j,i,j\}$ or
  $\{i,j,-i,-j\}$ and the template polyhedron in this limit is a
  cuboctahedron. Two pentagons at $A$ and $B$ fuse to become one
  square, and there are six octagons and twelve squares in total in
  this limiting case.
Other collinear cases do not make the third 
  side vanish though and pentagons will not fuse at all. In fact we 
  can use $T_3$ or $T_4$ introduced later to make these two base 
  vectors nonparallel.
 \end{enumerate} 
 When none of the sides of the scalene triangle vanishes, the
 corresponding octahedral fullerenes will be denoted as type IV.

\section{Index Symmetry}
In the previous section, we showed that an octahedral fullerene can be
constructed by cutting a fundamental polygon specified by a four-component index and
its category, $\{i,j,k,l\}_{X}$ and patching twenty-four replica of this
fundamental polygon onto a cantellated cube. We also pointed that this
correspondence is not one-to-one, but many-to-one, since there are some
symmetry relationships in this indexing scheme.  In other words, we
mean that there exist different indices $\{i,j,k,l\}_{X}$ that
correspond to the same molecular structure.  This
section is devoted to find a systematic way to eliminate all such redundancies
and fully characterize the nature of the index symmetry.

In the limiting cases of octahedral fullerenes which belong to the
types I to III, we only need one independent two-component vector to specify their
indices. It is obvious that the index transformation arising from the
geometric symmetry of graphene will lead to the same octahedral
fullerene. For instance, a $\pi/3$ rotation about point $O$ will
transform the index from $\{i,j,k, l\}_{X}$ to $\{-j,i+j,-l,k+l\}_{X}$
without altering the resulting octahedral fullerene. Therefore these
two indices correspond to the same molecular structure and should only
be counted once.  In fact, this applies to all twelve symmetry operations belonging to the point group 
$C_\text{6v}$ of graphene. Here, we ignore
symmetry operation $\sigma_h$ that lies in the plane of graphene
because it does not move any carbon atom at all. So, all indices
that can be related through these symmetry operations produce the
same octahedral fullerene.  This set of indices is called an
orbit in group theory\cite{fujita}. So to enumerate octahedral fullerene is equivalent to enumerate
different orbits of all possible indices. Indices belonging to the
same orbit correspond to the same octahedral fullerene.  In other
words, only one out of the set of indices comprising an
orbit is needed to represent an octahedral fullerene uniquely. In
these three limiting situations, we can restrict the indices with the
inequality, $i\ge j\ge 0\land i>0$ for type I, $k\ge l\ge0\land k>0$
for type II, and $i\ge j\ge 0\land i>0$ ($k=i$ and $l=j$) for type III
to remove all redundancies arising from the $C_\text{6v}$ symmetry
operations.

The situation for type IV octahedral fullerenes is more complicated.
In addition to the twelve symmetry operations from the
point group $C_{6v}$, there are three more symmetry operations,
$T_{2}$, $T_{3}$, and $T_{4}$ arising from different ways of dissecting
each of the three different kinds of faces of a cantellated cube into fundamental
polygons. For each dissection scheme, different squares or regular
triangles are drawn, and the square or triangular base vectors will
change respectively. Detailed description of these three symmetry
operations will be described later.  These extra symmetry operations
introduce redundancies which cannot be removed by introducing
inequalities of indices like the situations of types I to III.

Although the redundancies produced by these three $T$-type symmetry
operations cannot be removed by such index restrictions, the parts of
redundancies originating from the sixfold rotational symmetry of
graphene can be eliminated by introducing the canonical criterion,
$i>0\land j\ge0$.  This is because that these rotational operations
commute with the three $T$-type operations, {\it i.e.}
$\left[C_{6}^{n}, T_y\right]=0,$ where $y=2$, $3$ or $4$. Here, we do
not impose the restriction, $i\ge j$, to remove the redundancies
produced by the six mirror symmetries $M_{x}$. This will be discussed
with the $T_2$ symmetry in the next section.

\subsection{$T_{2}$ symmetry}
\begin{figure}
\includegraphics[width=15cm]{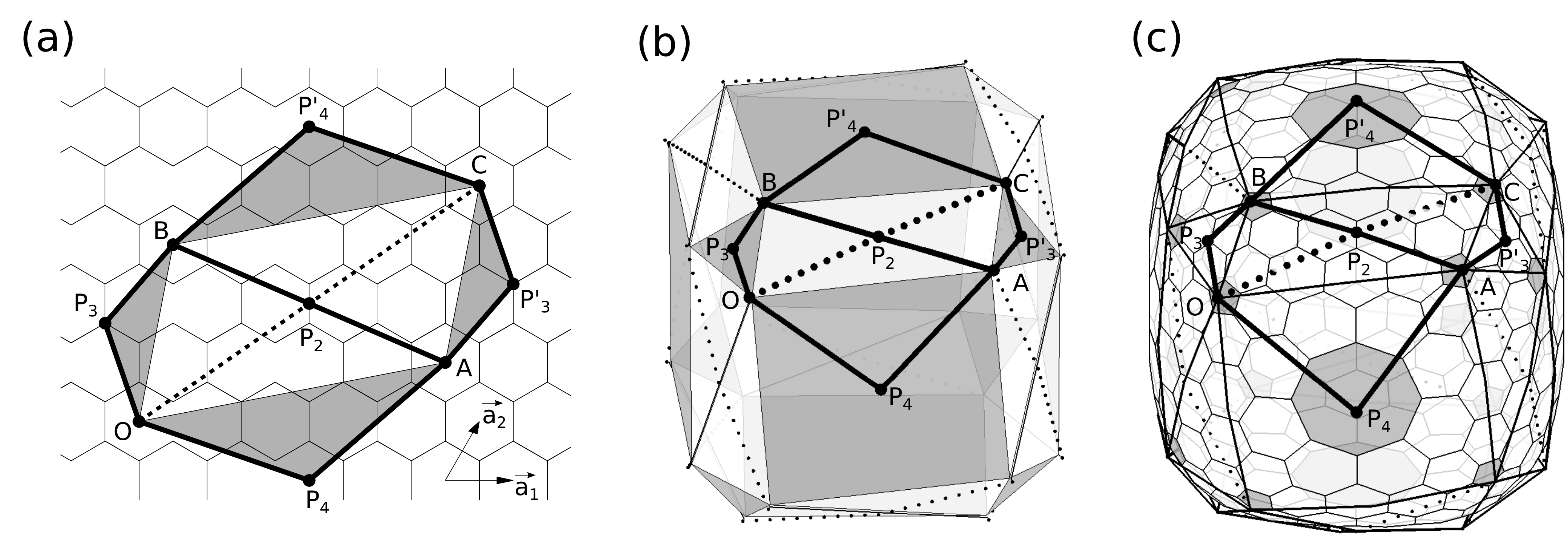}
\caption{The $T_{2}$ symmetry operation illustrated with the
  example $T_2\{4,1,-1,3\}_{\alpha}=\{4,1,1,-3\}_{\beta}$.  If we
  choose $\{\protect\overrightarrow{OA},
  \protect\overrightarrow{OB}\}_{\alpha}$ as the index, the
  corresponding fundamental polygon is $OP_{4}AP_{2}BP_{3}$.  On the
  other hand, if we choose the index
  $\{\protect\overrightarrow{BC},\protect\overrightarrow{BO}\}_{\beta}=\{\protect\overrightarrow{OA},
  -\protect\overrightarrow{OB}\}_{\beta}$, the fundamental polygon
  becomes $BP'_4CP_2OP'_3$.  These two fundamental polygons
  essentially give the same octahedral fullerene with different ways of
   dissecting the parallelogram.\label{fig:T2}}
\end{figure}

The symmetry operation, $T_{2}$, comes from the two different ways to
decompose a parallelogram as shown in Figure~\ref{fig:T2}.  The $T_2$
operation stands for performing a local $C_{2}$ operation which rotate
one of base vectors by $180^\circ$. Thus the index $\{i,j,-k,-l\}$
will generate the same octahedral fullerene with $\{i,j,k,l\}$.
We can define $T_2$ explicitly with the following matrix notation
\begin{align*}
  T_2:
  \begin{pmatrix}
    i\\j\\k\\l
  \end{pmatrix}_{X}
  \to
  \begin{pmatrix}
    i'\\j'\\k'\\l'
  \end{pmatrix}_{X'}=
  \begin{pmatrix}
    1&0&0&0\\0&1&0&0\\0&0&-1&0\\0&0&0&-1
  \end{pmatrix}
  \begin{pmatrix}
    i\\j\\k\\l
  \end{pmatrix}_{X},
\end{align*}
where $X\neq X'$.

Unlike usual matrix multiplications, we need to specify the category
of the index before and after $T_2$ transformation. Since $il-jk$
stands for the signed area enclosed by the parallelogram spanned by
these two vectors up to a positive factor, it is clear that under the
transformations, $T_2$ or $M_x$, the signed area changes sign and
hence the category. This is also true in the degenerate case.
Therefore, enumerating indices only in a single category can remove
redundancies produced by $T_2$ and $M_x$, but not those produced by
$M_xT_2=T_2M_x$.
% It is obvious that $T_2$ is its own inverse, $T_2^{-1}=T_2$.

\subsection{$T_{3}$ symmetry }
\begin{figure}
\includegraphics[width=15cm]{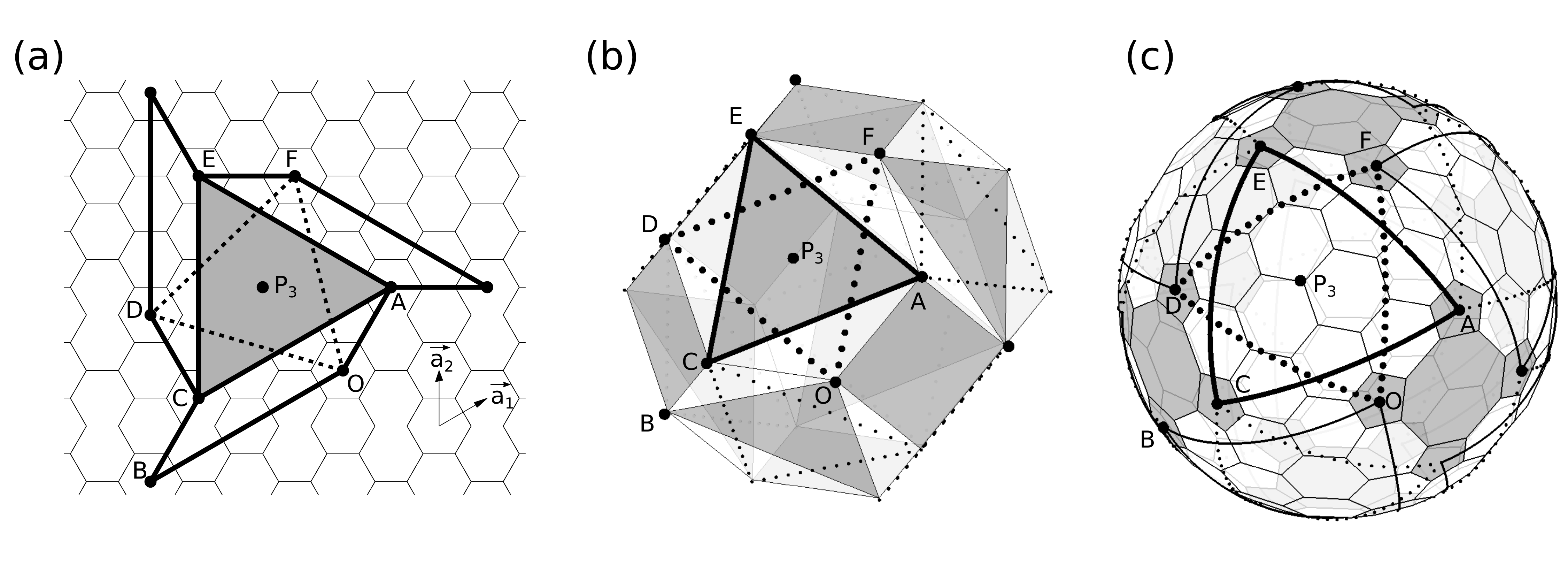} 
\caption{
  An illustration of $T_{3}$ symmetry.
  $T_{3}$ transform the partition 
  $\{\protect\overrightarrow{OA},\protect\overrightarrow{OB}\}_{\alpha}
  =\{1,1,-4,0\}_{\alpha}$
  to
  $\{\protect\overrightarrow{OA},\protect\overrightarrow{OF}\}_{\alpha}
  =\{1,1,-1,4\}_{\alpha}$, which can be also written as
  $T_{3,{\alpha}}\{1,1,-4,0\}_{\alpha}=\{1,1,-1,4\}_{\alpha}$. Similarly
  we have $T_3^{-1}\{1,1,-1,4\}_{\alpha}=\{1,1,-4,0\}_{\alpha}$.  
  \label{fig:T3}}
\end{figure}

The symmetry operation $T_{3}$ involves different ways of dissecting
the equilateral triangles of  the cantellated cube as shown in
Figure~\ref{fig:T3}. For instance, one possible choice of the two base
vectors for the scalene triangle is $\{\overrightarrow{OA},
\overrightarrow{OB}\}_{\alpha}$. However, there is another choice,
$\{\overrightarrow{OA},\overrightarrow{OF}\}_{\alpha}$, which produce
the same octahedral fullerene, but with a different way of dissecting
the triangles of the cantellated cube. The $T_3$ transformation only
changes the triangular base vectors.

Unlike $T_2$ and $M_x$, the $T_3$ transformation does not change the
category. Moreover, for the $T_{3,{\alpha}}$ transformation, which
operates on octahedral fullerenes belonging to the category $\alpha$,
we also need to impose an additional constraint on the domain
$i'l'-j'k'\ge 0 \Rightarrow - ik-jk-jl\ge i^2+ij+j^{2}$. The explicit
form of $T_{3,{\alpha}}$ can be
written as
\begin{align*}
  T_{3,{\alpha}}:
  \begin{pmatrix}
    i\\j\\k\\l
  \end{pmatrix}_{\alpha}
  \to
  \begin{pmatrix}
    i'\\j'\\k'\\l'
  \end{pmatrix}_{\alpha}=
  \begin{pmatrix}
    1&0&0&0\\0&1&0&0\\2&1&1&1\\-1&1&-1&0
  \end{pmatrix}
  \begin{pmatrix}
    i\\j\\k\\l
  \end{pmatrix}_{\alpha}.
\end{align*}
We may obtain $T_{3,{\beta}}$ easily by
$T_{3,{\beta}}=M_xT_{3,{\alpha}}M_x$ and its domain by similar method.
The inverse of $T_3$ transformation, namely $T_3^{-1}$, may be found
by the usual matrix inversion,
\begin{align*}
T_{3,{\alpha}}^{-1}:
\begin{pmatrix}
  i\\j\\k\\l
\end{pmatrix}_{\alpha}
\to
\begin{pmatrix}
  i'\\j'\\k'\\l'
\end{pmatrix}_{\alpha}=
\begin{pmatrix}
  1&0&0&0\\0&1&0&0\\-1&1&0&-1\\-1&-2&1&1
\end{pmatrix}
\begin{pmatrix}
  i\\j\\k\\l
\end{pmatrix}_{\alpha}
\end{align*}
Its domain can also be found by requiring that the category remains
unchanged, $i'l'-j'k'\ge 0 \Rightarrow ik+il+jk\ge i^2+ij+j^{2}$ and
so we have the identity,
$T_{3,{\beta}}^{-1}=M_xT_{3,{\alpha}}^{-1}M_x$.  In addition, as shown
in Figure~\ref{fig:T3}, the $T_3$ transformation always decrease
$|\theta|$ by more than $\pi/3$; while $T_3^{-1}$ always increase
$|\theta|$ by more than $\pi/3$.

\subsection{$T_{4}$ symmetry }
\begin{figure}
\includegraphics[width=15cm]{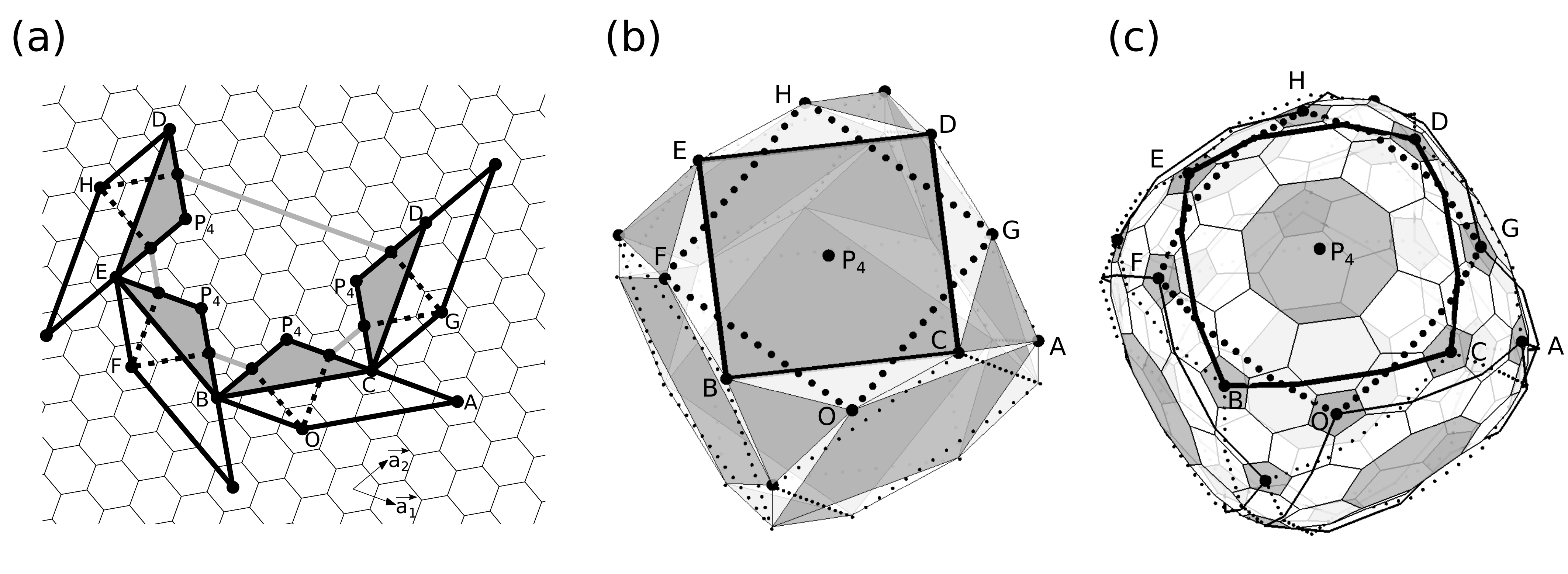}
\caption{An illustration of $T_{4}$ symmetry. $T_{4}$ transform the
  partition
  $\{\protect\overrightarrow{OA},\protect\overrightarrow{OB}\}_{\alpha}
  =\{2,2,-2,0\}_{\alpha}$ to
  $\{\protect\overrightarrow{OF},\protect\overrightarrow{OB}\}_{\alpha}
  =\{-4,2,-2,0\}_{\alpha}$, which can be also written as
  $T_{4,{\alpha}}\{2,2,-2,0\}_{\alpha}=\{-4,2,-2,0\}_{\alpha}$. Two
  points connected by a grey line should be patch into one point.
  Four shaded triangle in (a) will merge into square $BCDE$ in (b) and
  four $P_4$ points in (a) will become one $P_4$ in (b) after
  patching. Note that since $P_4$ always carries a topological charge, the vector $\protect\overrightarrow{OF}$ does not correspond to $(-4,2)$ in (a).\label{fig:T4}}
\end{figure}

Similar to the symmetry operations $T_2$ and $T_3$, the operation
$T_4$ involves different ways of assigning fundamental polygons on the
cantellated cube as shown in Figure~\ref{fig:T4}.  In this case, we
can see that two different fundamental polygons given by indices
$\{\overrightarrow{OA},\overrightarrow{OB}\}_{\alpha}$ and
$\{\overrightarrow{OF},\overrightarrow{OB}\}_{\alpha}$ are essentially
equivalent in constructing an octahedral fullerene. The transformation
$T_4$ does not change the category just like the transformation
$T_3$. We can interchange $T_{4,{\alpha}}$ and $T_{4,{\beta}}$ by
sandwiching them between the mirror transformation $M_x$.  On the
other hand, in contrast to the transformation $T_3$, $T_4$ changes
the square base vector only. Therefore, both $T_3$ and $T_4$ will decrease
$|\theta|$ by more than $\pi/3$. In other words, the square base
vector will be rotated by more than $\pi/3$ and will not satisfy the
canonical criterion $i>0\land j\ge 0$ any longer. However the whole
index can be rotated back to satisfy the canonical criterion again
whenever necessary.

The explicit form for the symmetry operation $T_{4,{\alpha}} $ can be
written as
\begin{align*}
  T_{4,{\alpha}} :
  \begin{pmatrix}
    i\\j\\k\\l
  \end{pmatrix}_{\alpha}
  \to
  \begin{pmatrix}
    i'\\j'\\k'\\l'
  \end{pmatrix}_{\alpha}=
  \begin{pmatrix}
    0&-1&1&-1\\1&1&1&2\\0&0&1&0\\0&0&0&1
  \end{pmatrix}
  \begin{pmatrix}
    i\\j\\k\\l
  \end{pmatrix}_{\alpha}.
\end{align*}
Again, an constraint on domain $-ik-jk-jl\ge k^2+kl+l^2$ is necessary
to ensure that the category stays unchanged.  The inverse
$T_{4,{\alpha}} ^{-1}$ can be defined as follows
\begin{align*}
  T_{4,{\alpha}} ^{-1}:
  \begin{pmatrix}
    i\\j\\k\\l
  \end{pmatrix}_{\alpha}
  \to
  \begin{pmatrix}
    i'\\j'\\k'\\l'
  \end{pmatrix}_{\alpha}=
  \begin{pmatrix}
    1&1&-2&-1\\-1&0&1&-1\\0&0&1&0\\0&0&0&1
  \end{pmatrix}
  \begin{pmatrix}
    i\\j\\k\\l
  \end{pmatrix}_{\alpha},
\end{align*}
and the constraint on the domain is $ik+il+jl\ge k^2+kl+l^2$. In 
summary, $C_6^n$, $T_3$, $T_4$ and their inverses do not change the
categories, but $M_x$ and $T_2$ do.

Although $T$-type symmetry operations are defined for type IV
octahedral fullerenes, they can also be applied to three limiting
cases. When $T$-type symmetry operations are applied to type I and
type II octahedral fullerenes, they reduce to the geometric rotation
$C_6^n$. And when they are applied to type III octahedral fullerenes,
we have following identities,
\begin{align*}
  T_2\{i,j,i,j\}_{X}&=\{i,j,-i,-j\}_{X'}\quad(X\neq X')\\
  T_3^{-1}\{i,j,i,j\}_{X}&=\{i,j,-i,-j\}_{X}\\
  C_6^3T_4^{-1}\{i,j,i,j\}_{X}&=\{i,j,-i,-j\}_{X}.
\end{align*}
These formulae will give a torus-like orbit. The details for the
enumeration of these orbits are included in supporting information.

\section{Conclusion}
\label{sec:conclusion}

In conclusion, we have developed a systematic cut-and-patch method to
generate arbitrary fullerenes belonging to the octahedral point
group. A unique four-component vector satisfying certain constraints and
symmetry rules can be used to specify these octahedral
fullerenes. This work on the octahedral fullerenes fits in the final
piece of the jigsaw puzzle of all possible high symmetry caged
fullerenes based on Platonic solids.  Further investigation on the
stability, elastic properties and electronic structures of these
octahedral fullerenes and the possibility of using them to build
periodic carbon Schwarzites are currently undergoing in our
group\cite{tomanek,jin:2010x}.

Finally, we also want to point out two observations: the ``Brazuca''
ball used in the World Cup is close to a very round octahedral sphere,
while the fullerenes discussed in this paper are still far from a
round sphere. The explanation for the first observation is given in a
more general context by Delp and Thurston in a paper about the
connection between clothing design and mathematics in the Bridges
meeting three years ago.\cite{thurston} The most important factor that
makes it possible to wrap six clover-shaped panels used in the ``Brazuca''
around a sphere smoothly is that the curved seams created by these
interlocked 4-long-arms panels are quite evenly distributed on the
sphere. Readers interested in this problem should go to that
paper for details. The observation on the shape of octahedral
fullerenes is also interesting. All of the three-dimensional
geometries shown in this paper are obtained through their topological
coordinates derived from the lowest three eigenvectors with single
nodes by diagonalizing the corresponding adjacency
matrices\cite{Fowler07}. Further investigations to rationalize how the
distribution of the non-hexagons affects the shapes of octahedral
fullerenes in order to obtain a round nanoscale ``Brazuca'' ball based
on either elastic theory or quantum chemical calculations should be
worth pursuing in the future.\cite{Fowler07,tomanek,siber,nelson}
\section*{acknowledgements} 

The research was supported by the Ministry of Science and Technology, Taiwan.
B.-Y. Jin thanks Center for Quantum Science and Engineering, and
Center of Theoretical Sciences of National Taiwan University for
partial financial supports. We also wish to thank Chern Chuang and
Prof. Yuan-Chung Cheng for
useful discussions and comments on this paper.

\end{document}